\documentclass[12pt,preprint]{aastex}

\def\kms{{\>\rm km\>s^{-1}}}

\shorttitle{Spectroscopic Binaries in PNe}
\shortauthors{De Marco et al.}

\begin{document}

\title{Indications of a Large Fraction of Spectroscopic Binaries Among 
Nuclei of Planetary Nebulae}

\author{Orsola De Marco\altaffilmark{1},
Howard E. Bond\altaffilmark{2},
Dianne Harmer\altaffilmark{3}, and
Andrew J. Fleming\altaffilmark{4}
}

\altaffiltext{1}
{Department of Astrophysics, American Museum of Natural History,
Central Park West at 79th St., New York, NY 10024; orsola@amnh.org}

\altaffiltext{2}
{Space Telescope Science Institute, 3700 San
Martin Drive, Baltimore, MD 21218; bond@stsci.edu}

\altaffiltext{3}
{Kitt Peak National Observatory, National Optical
Astronomy Observatories, P.O. Box 26732, Tucson, AZ 85719; diharmer@noao.edu}

\altaffiltext{4}
{REU fellow, summer 2003, American Museum of Natural History; Department
of Physics, Michigan Technological University, 1400 Townsend Drive, Houghton,
MI 49931; ajflemin@mtu.edu}

\begin{abstract}

\looseness=-1
Previous work indicates that about 10\% of planetary-nebula nuclei 
(PNNi) are photometrically variable
short-period binaries with periods of hours to a few days. These systems have 
most likely descended from common-envelope (CE) interactions in initially much
wider binaries. Population-synthesis studies suggest that these very close
pairs could be the short-period tail of a much larger post-CE binary population
with periods of up to a few months. We have initiated a radial-velocity (RV)
survey of PNNi with the WIYN 3.5-m telescope and Hydra
spectrograph, which is aimed at discovering these intermediate-period binaries.
We present initial results showing that 10 out of 11 well-observed PNNi have
variable RVs, suggesting that a significant binary population may be present. 
However, further observations are required because we have as yet been unable
to fit our sparse measurements with definite orbital periods, and because some
of the RV variability might be due to variations in the stellar winds of some
of our PNNi.

\end{abstract}


\section{Introduction}

There is reason to think that many of the central stars in
planetary nebulae (PNe) may be close binaries.  The evidence leading to this
suggestion includes the following:

(1)~Photometric monitoring has revealed that 
about 10\% of planetary-nebula nuclei (PNNi) are very close binaries, with
periods of a few hours up to a few days (e.g., Bond \& Livio 1990; Bond 
2000). These close binaries in PNe provide direct evidence for the
occurrence of a common-envelope (CE) interaction (Paczynski 1976;
Sandquist et~al.\ 1998), 
in which one component of a binary evolves to giant dimensions, and
then engulfs a main-sequence companion; the ensuing spiral-down of the orbit
ejects the CE and exposes the hot core of the red giant, leaving a close binary
inside a photoionized PN.

\def\ace{\alpha_{\rm CE}}

(2)~Theoretical studies of the evolution of binary populations
(e.g. Yungelson et~al.\ 1993, Han
et~al.\ 1995), predict the orbital period distribution of binary
stars inside PNe to be a strong function of the
efficiency, $\ace$, with which the orbital energy of the original
system goes into ejecting material from the CE\null.  
However, a recent study (O'Brien
et~al.\ 2001) of the post-CE eclipsing binary V471~Tauri
indicates that $\ace\approx0.1$, for this one object.  Fig.~3b
of Yungelson et~al.\ then predicts, if this value of $\ace$ is 
applicable to most CE interactions, that the orbital periods of binaries in PNe
should be distributed roughly evenly over the range 
0.3--30~days.  Since 10\% of PNNi are already known to be binaries lying in the
short-period tail that is detectable photometrically, these results suggest
that a large fraction of PNNi could be longer-period binaries.  

(3)~A large majority of PNe have highly non-spherical shapes, including
numerous extreme cases of strongly bipolar or axisymmetric morphologies (e.g.,
Zuckerman \& Aller 1986; Soker 1997). The simplest
explanation for these shapes would be that most PNe have been ejected through
CE interactions, or that the PN ejection process has at least been affected by
other phenomena directly related to the presence of a companion star (e.g.,
tidal spinup and/or dynamo generation of magnetic fields).

The strongest empirical test of this hypothesis of a large binary fraction
would be to search for the
expected population of binary PNNi with periods too long to be detected from
photometric variability, but detectable through radial-velocity (RV) variations.
Knowledge of the overall period distribution of binary PNNi would provide
strong constraints on (a)~the binary properties of the parent AGB population,
and (b)~the typical value of $\ace$, a quantity needed to predict the
properties of other post-CE systems, including cataclysmic variables, low-mass
X-ray binaries, and the progenitors of Type~Ia supernovae.

In this Letter we report initial results of a RV survey of PNNi, designed to
detect variability on timescales of a few days up to a few months, with
velocity amplitudes down to a few $\kms$.

\section{Target Selection and Observations}

Our spectroscopic data were taken at the 3.5-m WIYN telescope at Kitt Peak
National Observatory between 2002 August and 2003 September.  It is well known
(e.g., Kennicutt, Freedman, \& Mould 1995) that  for optimum sampling of
unknown periods, the observations should be spaced over an interval at least as
long as the longest expected period, with the intervals between successive
observations increasing according to a power series. 

Since we wished to search for periods of a few days up to about two months,
this scheduling requirement dictated use of a spectrograph that is easily
available on the telescope without major instrument changes, making runs as
short as one night possible. The Hydra fiber-optics bench-mounted spectrograph
(Barden \& Armandroff 1995), which can be put into use very quickly on the WIYN
telescope  by rotating the Nasmyth tertiary mirror, was thus ideal for meeting
our scheduling needs.  We were awarded 8 nights each in three successive
semesters, with each semester's individual nights spaced nearly ideally as
outlined above.   Unfortunately, however, we encountered unusually bad
weather---not to mention a nearby forest fire---during all three semesters, so
that the actual sampling that we achieved was very far from ideal.

We used the Hydra spectrograph with fibers that project to a  $2\farcs0$
diameter on the sky; this diameter is near optimal for collecting most of the
starlight in typical seeing, while giving little degradation of the spectral
resolution, and permitting minimal contamination from the surrounding
nebulae.   A 1200 groove~mm$^{-1}$ grating, combined with the ``Red Bench''
camera and the T2KC CCD detector, provided a dispersion of
0.33~\AA~pixel$^{-1}$ and a wavelength coverage of 4050--4730~\AA\null.  The
FWHM of comparison lines was typically 0.60~\AA\null.  We generally obtained a
Cu-Ar comparison lamp exposure after every three stellar exposures, which was
adequate because of the high stability of the bench spectrograph.

Our goal was to conduct a RV survey of a representative selection of PNNi.
However, as is well known, many PNNi have significant stellar winds, raising
the danger that short-timescale variations in mass-loss rate could give rise to
spurious apparent velocity variations.  (Note, for example, that virtually all
PNNi with obvious P~Cygni profiles in the ultraviolet are observed to vary
photometrically on short timescales---Bond \& Ciardullo 1989; Handler et~al.\
1997.)  In order to maximize the probability of measuring center-of-mass
motions of the stars, we drew up a list of targets likely to have low mass-loss
rates and optical spectra dominated by photospheric absorption lines.  These
included stars classified as type~O (e.g., by Heap 1977), sdO, B, or sdB by
various authors.  In order to minimize exposure times, we favored stars
brighter than about 14th  visual magnitude.  Several candidate stars were
observed, but found to have few or no usable absorption lines, and were thus
dropped from our program.  At this writing, we have obtained useful spectra of
11 PNNi on at least 4, and up to 16, different nights.

In addition to the PNNi, we obtained spectra on most nights of the well-known
sdO spectrophotometric standard star BD~$+28^\circ$4211.  This star has a
spectrum similar to those of some of our program stars, but is not known to be
surrounded by a PN\footnote{As noted by Zanin \& Weinberger (1997), the Palomar
Sky Survey shows some diffuse nebulosity with an angular extent of several
degrees in the vicinity of BD~$+28^\circ$4211. If actually associated with the
star, this nebulosity would have a physical scale much larger than any other
known PN.}.  To the best of our knowledge this star has not been reported to
have a variable RV, and thus it should serve to monitor the stability of our
velocity measurements for the program stars.

\section{Data Reduction and Analysis}

We reduced our spectra with NOAO IRAF\footnote{IRAF is distributed by the
National Optical Astronomy Observatories, which are operated by the Association
of Universities for Research in Astronomy, Inc., under cooperative agreement
with the National Science Foundation.} software.  The {\it dohydra\/} task was
used to flat-field, extract, and wavelength-calibrate  the spectra. Excellent
fits to the wavelength dispersion curve were obtained, with RMS scatter of
about 0.03~\AA\null. 

We measured RV shifts between normalized pairs of spectra using the
cross-correlation technique implemented by the IRAF task {\it fxcor}.  For each
star in our sample, we initially selected the spectrum with the highest
signal-to-noise ratio (SNR) as the template, and cross-correlated it with the
other spectra taken at different epochs. We then used these initial RV 
estimates to shift each spectrum to the velocity of the template, and created
a  weighted average of all of them. This averaged spectrum, with typical $\rm
SNR\ge100$, was then used as the final template to determine the relative
RV shifts between it and each of the spectra. In the present work, we have not
attempted to place the RVs on an absolute velocity scale. 

The {\it fxcor\/} task calculates the velocity shift between two spectra by
fitting the correlation with a user-selected function. Most of our PNNi
have a variety of line profiles, including pressure-broadened Balmer lines as
well as narrower lines from metallic species.  A parabola was finally
adopted as the best fitting function.

We regard the errors in our relative velocities as arising from two sources.
One is the random velocity scatter due to errors in the dispersion fit,
mechanical drifts in the spectrograph, and other less easily specified effects.
To estimate this scatter, we measured the relative velocities of the strong
nebular [\ion{O}{3}] 4363~\AA\ emission line of NGC~6891, based on 17 spectra.
Such a sharp nebular line is expected to have a constant RV\null.  The RMS
scatter of the measured relative velocities was found to be $2.1\kms$, which we
take as a measure of the random velocity errors affecting all of our
measurements (note that this is closely similar to the velocity scatter
expected from the 0.03~\AA\ $\simeq$ 2.0~$\kms$ dispersion of our
comparison-line fits).  

In addition, there is a contribution from the errors of the individual
absorption-line cross-correlation fits due to effects such as photon noise and
the shapes and widths  of the absorption-line profiles.  The {\it fxcor\/} task
reports a random error for each velocity shift based on the error in fitting a
parabola to the cross-correlation function.   For a sharp, strong emission line
such as [\ion{O}{3}] 4363~\AA\ in NGC~6891,  the error reported by {\it
fxcor\/} is negligible ($<$$0.2\kms$) compared to the night-to-night scatter. 
For the absorption-line RVs of our program stars, we found that the {\it
fxcor\/} errors are also less than the random night-to-night error estimated
above, except for spectra with low SNR, or with especially broad or shallow
absorption lines.  In our analysis below, we combine in quadrature the random
error estimate of $2.1\kms$ with the random error from {\it fxcor}, in order to
estimate the overall error of each RV measurement.


The stellar absorption  lines most commonly encountered in PNNi in our spectral
range are H$\delta$ and H$\gamma$ (often contaminated by nebular emission), as
well as lines of \ion{He}{1}, \ion{He}{2}, \ion{C}{3}, \ion{C}{4}, \ion{N}{3},
and \ion{N}{5}. When determining the cross correlation using stellar absorption
lines, we selected spectral windows so as to exclude the nebular emission lines
in those objects with strong superposed emission. 

\section{Results and Discussion}

In Table~1 we present the list of stars for which we have obtained at least 4
good (continuum $\rm SNR\ge20$) spectroscopic observations. In columns (5),
(6), and (7), we list the number of times each star was observed, $n$, the
standard deviation, $\sigma$, of the ensemble of relative RV measurements, and
the average error of an individual RV measurement.  Each star was observed once
per night, on as many nights as the weather and sidereal time permitted.

The question whether a collection of RV measurements is consistent with the
hypothesis of a constant velocity is a classical problem, discussed in detail 
for example by Trumpler \& Weaver (1953).  These authors recommend a standard
$\chi^2$ calculation, from which the reduced value, $\chi^2/(n-1)$ 
(given in Table~1, column [8]), leads to an
estimate of the probability that the star's RV is variable 
(column [9]; e.g., Press et~al.\
1986).  

Table~1 shows the startling conclusion that 10 out of the 11 PNNi that we have
observed have variable velocities.  Reassuringly, our control star,
BD~$+28^\circ$4211, has a constant RV according to our statistical test (i.e.,
it has only a 24\% probability of variability).  (As Table~1 shows, the
individual RV errors for  BD~$+28^\circ$4211 are relatively large, averaging
$3.3\kms$.  This is due to the relatively high gravity of this star, which
gives it broader lines than our typical PNNi; hence the cross-correlation
function is broader, with an attendant larger error in the velocity shifts.)

Are these RV variations due to motion in a binary system, or to some other
phenomenon (e.g., stellar-wind variations that modulate absorption-line
profiles in a way that mimics velocity variability)?  Detection of a clearly
periodic velocity variation would provide strong support for binary motion.  
Unfortunately, due to the poor weather that we encountered, we have only a
handful of observations of each star spaced over more than a one-year
interval.  Most of our stars appear to show RV variations on timescales as
short as 1~day, suggesting that their periods, if they are binaries, are
relatively short. Thus our sampling is extremely non-optimal for period
searching. We have used the Lafler \& Kinman (1965) periodogram to search for
periodic signals, but due to the severe aliasing it is not possible to find a
definite period for any of our stars; a very wide variety of periods can yield
more-or-less smooth RV curves, but none of them stand out clearly.

In order to investigate further for the presence of winds, we examined UV
spectra of our targets obtained with the {\it International Ultraviolet
Explorer\/} ({\it IUE\/}) satellite, using the data archive at the Space
Telescope Science Institute\footnote{http://archive.stsci.edu/iue/index.html}. 
All 12 of our targets, including BD~$+28^\circ$4211, have been observed with
{\it IUE\/}, but only five of them (IC~4593, NGC~6210, NGC~6891, A~78, and
Sa~4-1) show P~Cygni profiles in the UV\null.  However, most optical absorption
lines are much less affected by winds than the  UV resonance lines. Moreover,
asymmetric profiles would typically be introduced for lines formed in a wind,
but we did not observe any unusual profiles in the lines measured for RV (note
that {\it fxcor\/} directly displays the correlation function, which has the
same shape as the typical stellar absorption lines), nor any
obvious time variations in the line profiles. Thus, although we cannot
absolutely exclude that the RVs  of these five targets are affected by wind
phenomena, we believe it to be unlikely; and in any case, we find it hard to
believe that the variations of the other five  variables are due to such a
phenomenon. 

We thus have a tantalizing suggestion that the binary fraction
among PNNi may be very high.  As noted above, photometric monitoring has
indicated that about 10\% of randomly selected PNNi are short-period (hours to
a few days) binaries.  The RV results presented in this paper now suggest that
an even larger fraction (which may well exceed 50\%, and may approach 100\%) of
random PNNi may be spectroscopic binaries of intermediate periods. Our results
appear to be consistent with those presented recently by Pollacco (2003), who
finds, in a somewhat less sensitive RV survey, that 57\% of a sample of 23 PNNi
have variable RVs. 

We plan to continue our monitoring of northern-hemisphere targets, and we are
currently analyzing RV data for about 30 southern-hemisphere PNNi. Such
observations will greatly clarify the situation. At present, it has become
increasingly plausible that {\it binary-star ejection is a major formation
channel for planetary nebulae}.

\bigbreak

\centerline{\bf APPENDIX}

Comments on individual objects:

{\bf PHL~932} is almost unique among PNNi in being classified as a sdB star. 
Its effective temperature and surface gravity (Napiwotzki 1999)  suggest it
could be a post-red giant in a binary system that underwent a CE episode
(Mendez et~al.\ 1988; Iben \& Tutukov 1993). No RV variations greater than
$2\kms$ were detected over an interval of 6~days by Wade (2001).  Our claim of
RV variability rests largely on one outlying velocity measurement, differing
from the mean of the remaining 8 observations by $-8.9\pm2.5\,\kms$; this
suggests that, if the star is a spectroscopic binary, its eccentricity is high,
which would be surprising for a post-CE system.  

{\bf BD +33$^\circ$2642} is an extremely well-observed spectrophotometric
standard star, so it is a surprise that its surrounding faint PN was not
discovered until 10~years ago (Napiwotzki 1993). Our RV measurements show a
range of $15.3\kms$, well in excess of the small errors for this bright star
with many usable photospheric absorption lines.

{\bf IC 4593, NGC 6210, NGC~6891}. Our measurements of these three rather
similar objects rely on absorption lines of \ion{He}{2} at 4200 and 4541~\AA,
as well as the wings of the Balmer lines (the cores are contaminated with
nebular emission).  We find definite RV variability in all three, with total
ranges of 36.4,  16.7, and $15.6\kms$, respectively.   However, the nucleus of
IC~4593 is variable photometrically on a timescale of hours (Bond \& Ciardullo
1989), has P~Cygni profiles in the UV, and shows short-timescale variations in
its optical  (Mendez, Herrero, \& Manchado 1990) and UV (Patriarchi \&
Perinotto 1995) emission features.  Mendez et~al.\ also reported optical
absorption-line RV variations in IC~4593 similar to those we report here. The
other two PNNi likewise show P~Cygni profiles in the UV\null.   We therefore
cannot completely rule out that the optical RV variability that we measured is
due to variations in the stellar-wind mass loss.

{\bf IRAS 19127+1717} was suspected of binarity by Whitelock \& Menzies (1986),
because the contrast between the high excitation of the PN with the relatively
cool temperature of the B9 central star suggested that the ionization source is
an optically inconspicuous hot companion.  We find strong evidence for RV
variability, with a total range of $33.0\kms$.

{\bf LS IV $-$12$^\circ$111}, originally discovered in the Case-Hamburg survey
for luminous early-type stars, was re-classified as a post-AGB star embedded in
a young PN by Conlon et~al.\ (1993).  Arkhipova et~al.\ (2002) have reported
photometric variations, and Ryans et~al.\ (2003) suspected RV variability on
the basis of 3 velocity measurements.  We confirm definite RV variability, with
a total range of $39.0\kms$.

{\bf M 1-77, M 2-54} have both been reported to be photometric variables on
timescales of hours (Handler 1995, 1999). We find definite RV variability for
both PNNi (total ranges of 32.8 and $49.0\kms$, respectively).  Although the
photometric variability might suggest a connection with the wind variability
discussed above for IC~4593 and similar objects, our spectra show a wealth of
sharp absorption lines, and {\it IUE\/} spectra show no P~Cygni profiles.  We
thus believe there is a high probability that both stars are spectroscopic
binaries, although the non-periodic  photometric variability would then be left
unexplained.

{\bf A~78} is a prototypical born-again PNN (Iben et~al.\ 1983; Jacoby \& Ford
1983). Like IC~4593, NGC 6210, and NGC~6891, it has pronounced P~Cygni profiles
in the UV, so the RV variations that we find (total range $23.1\kms$) might
conceivably be attributable to wind variability.

{\bf Sa~4-1} has a high-velocity wind seen as blue-shifted absorption in 
high-dispersion {\it IUE\/} spectra, with relatively weak emission near zero
velocity (Feibelman \& Bruhweiler 1989), putting it at risk of wind
variability. Our 4 spectra, however, show only marginal evidence for RV
variation.

\acknowledgments

We are thankful for the terrific effort of the WIYN/NOAO team in supporting our
project, in particular WIYN Observatory Director George Jacoby, telescope
operators Gene McDougall, George Will, Doug Williams, and Hillary Mathis,   and
instrument specialist Chuck Corson. We thank John Glaspey for carrying out the
difficult task of meeting our exacting telescope scheduling requirements. Don
Pollacco provided useful information in advance of publication. OD is grateful
to Janet Jeppson Asimov for financial support.

\begingroup

\catcode`?=\active
\def?{\phantom{1}}

\begin{deluxetable}{llclccccc}
\tabletypesize{\small}
\tablecaption{Target list and Radial-Velocity Measurements }
\tablewidth{0pt}
\tablehead{
\colhead{Star} & \colhead{PN G} & \colhead{$V$} & \colhead{Sp.}  & \colhead{$n$} &
  \colhead {$\sigma$} & \colhead {Error of 1 obs} &  \colhead{$\chi^2$} & \colhead{$P$} \\
\colhead{    } & \colhead{}     & \colhead{mag} & \colhead{Type} & \colhead{  }  &
  \colhead {($\kms$)} & \colhead {($\kms$)} & \colhead{} & \colhead{var}}
\startdata
\multicolumn{9}{c}{ }\\
\multicolumn{9}{c}{RV variables}\\
\multicolumn{9}{c}{ }\\
PHL 932              & 125.9$-$47.0 & 12.1 & hg O (H)  & ?9 & ?3.8 & 2.6 & ?2.28 & 0.98 \\
BD $+33^\circ$2642   & 052.7+50.7   & 10.8 & B2 IVp    & 14 & ?3.7 & 2.3 & ?2.62 & 1.00 \\
IC 4593              & 025.3+40.8   & 11.2 & O5f (H)   & ?8 & 11.9 & 3.0 & 18.19 & 1.00 \\
NGC 6210             & 043.1+37.7   & 12.7 & O (H)     & ?6 & ?5.8 & 2.4 & ?6.67 & 1.00 \\
IRAS 19127+1717      & 051.0+02.8   & 13.4 & B9 V      & 12 & ?9.5 & 3.1 & ?9.94 & 1.00 \\
LS IV $-12^\circ$111 & 029.1$-$21.2 & 11.4 & B1 Ibe    & 15 & 12.1 & 2.3 & 28.78 & 1.00 \\
NGC 6891             & 054.1$-$12.1 & 12.4 & Of (H)    & 16 & ?4.6 & 2.6 & ?3.16 & 1.00 \\
M 1-77               & 089.3$-$02.2 & 12.1 & OB        & 15 & ?9.5 & 2.5 & 15.43 & 1.00 \\
A 78                 & 081.2$-$14.9 & 13.2 & Of/WR (C) & 11 & ?5.1 & 2.6 & ?3.99 & 1.00 \\
M 2-54               & 104.8$-$06.7 & 12.1 & B         & 15 & 11.8 & 2.6 & 21.30 & 1.00 \\
\multicolumn{9}{c}{ }\\
\multicolumn{9}{c}{Possible RV variable}\\
\multicolumn{9}{c}{ }\\
Sa 4-1               & 075.7+35.8   & 14.3 & O (H)     & ?4 & ?2.4 & 2.4 & ?1.25 & 0.71 \\  
\multicolumn{9}{c}{ }\\
\multicolumn{9}{c}{Control star}\\
\multicolumn{9}{c}{ }\\
BD $+28^\circ$4211   & \hfil$\dots$ & 10.5 & sdO       & 14 & ?2.9 & 3.3 & ?0.70 & 0.24 \\
\enddata
\tablecomments{Cols.~1 and 2: target name and designation in PN~G nomenclature;
cols.~3 and 4: $V$ magnitude and spectral type, from Acker et~al.\ (1992) or
SIMBAD database; col.~5: number of spectroscopic observations; col.~6: weighted
RMS scatter of our RV measurements; col.~7: weighted mean error of one RV
observation; col.~8: $\chi^2$ per degree of freedom; col.~9: probability that
the star's RV is variable.}
\end{deluxetable}

\endgroup

\end{document}